\documentclass[a4paper, 12pt, oneside]{article}		
\usepackage{amsfonts, 
amsmath, 
amssymb, 
theorem, 
amsfonts,			
sectsty				
}

\newtheorem{thm}{Theorem}
\newtheorem{corl}[thm]{Corollary}
\newtheorem{lma}[thm]{Lemma}
\newtheorem{prop}[thm]{Proposition}
{\theorembodyfont{\rmfamily}\newtheorem{defn}[thm]{Definition}
{\theorembodyfont{\rmfamily}\newtheorem{ex}[thm]{Example}

\newcommand{\ii}{\mathrm{i}} 
\newcommand{\bC}{\mathbb{C}} 
\newcommand{\bR}{\mathbb{R}} 

\newcommand{\bZ}{\mathbb{Z}} 
 
\newcommand{\bT}{\mathbb{T}}
\newcommand{\bdd}{\mathbb{\mathrm{d} }}
\newcommand{\dd}{\mathrm{d}}

\newcommand{\cA}{\mathcal{A}} 
\newcommand{\cC}{\mathcal{C}}
\newcommand{\cS}{\mathcal{S}} 
\newcommand{\cB}{\mathcal{B}}

\newcommand{\cH}{\mathcal{H}}

\newcommand{\cL}{\mathcal{L}}

\newcommand{\kast} { {[\ast]} }
\newcommand{\Dom} {\mathrm{Dom} }
\newcommand{\End} {\mathrm{End} }
\newcommand{\id} {\mathrm{id} }
\newcommand{\Sp} {\mathrm{sp} }

\newcommand{\Dirac}{/ \!\!\!\! D}

\newcommand{\trace}{\mathrm{Tr}}

\newcommand{\sign}{\mathrm{sign}}

\newcommand{\intC}{\int \!\!\!\!\!\! - \mbox{ }}
\newcommand{\triple}{(\cA, \cH, D)}
\newcommand{\tensor[1]}{\mbox{ }\overline{\otimes}_{#1} \mbox{ }}

\newcommand{\bd}{\begin{displaymath} }
\newcommand{\ed}{\end{displaymath} }
\newcommand{\be}{\begin{equation}}
\newcommand{\ee}{\end{equation}}
\newcommand{\bea}{\begin{eqnarray}}
\newcommand{\eea}{\end{eqnarray}}

\newcommand{\rightbox}{\begin{flushright}$\Box$\end{flushright} }

\newcommand{\mb}{\mbox{ }}

\newcommand{\dcite}[1]{\cite{#1}}

\begin{document}
\title{The noncommutative Lorentzian cylinder\\as an isospectral deformation}
\author{W.D. van Suijlekom\thanks{Current address: Scuola Internazionale Superiore di Studi Avanzati, Via Beirut 2-4, 34014 Trieste, Italy. E-mail: \tt{wdvslkom@sissa.it}}
\\[3mm]
Korteweg-de Vries Institute for Mathematics\\ University of Amsterdam\\ Plantage Muidergracht 24\\1018 TV Amsterdam\\The Netherlands }
\maketitle
\begin{abstract}
We present a new example of a finite-dimensional noncommutative manifold, namely the noncommutative cylinder. It is obtained by isospectral deformation of the canonical triple associated to the Euclidean cylinder. We discuss Connes' character formula for the cylinder. 

In the second part, we discuss noncommutative Lorentzian manifolds. Here, the definition of spectral triples involves Krein spaces and operators on Krein spaces. A central role is played by the admissible fundamental symmetries on the Krein space of square integrable sections of a spin bundle over a Lorentzian manifold. Finally, we discuss isospectral deformation of the Lorentzian cylinder and determine all admissible fundamental symmetries of the noncommutative cylinder. 
\end{abstract}
\pagebreak

\sectionfont{\large}
\section{Introduction}
Strict deformation quantization \cite{80,90,landsman} provides a powerful mathematical tool to describe the notion of quantization in physics. The central object here is a family of $C^\ast$-algebras $\{ \cA_\hbar \}$, parametrized by some real number $\hbar$. Recall that a $C^\ast$-algebra $\cA$ is a norm-closed $*$-algebra where the norm satisfies
\bd
\| a^\ast a \| = \|a \|^2, \quad (a \in \cA).
\ed
In physics, the commutative algebra of functions on a phase-space describes a classical theory. We denote this algebra by $\cA_0$. A quantum mechanical theory at value $\hbar$ of Planck's constant, on the other hand, is described by a noncommutative algebra of operators on a Hilbert space, denoted by $\cA_\hbar$ ($\hbar \neq 0$). As we will see, the family $\{ \cA_\hbar\}$ is a strict deformation quantization if it satisfies certain axioms. 

A good example is the noncommutative torus. It is obtained via deformation quantization of the algebra of functions on the torus $\bT^d$ \cite{80,83}. The noncommutative tori play a role in string theory and M(atrix) theory \cite{71}.

In this article, we will discuss a third example: the noncommutative cylinder. It is defined along the same lines by deformation quantization of the cylinder. In string theory, the cylinder is quite a natural object. There, space-time is a manifold of dimension higher than four. This dimension follows from certain consistency conditions of the theory (see Polchinski \dcite{polchinski1}). For example, the superstring can only be defined in a ten-dimensional background, say $\bR^{10}$. It is usually toroidally compactified to $\bR^4 \times \bT^6$ in order for the theory to make sense. This means that six dimensions are rolled up to the $6$-torus $\bT^6$. As Seiberg and Witten argued in \dcite{3}, the effective action of open strings in the presence of a constant magnetic field in the background is described by making space-time noncommutative. In order to describe this noncommutative background, one needs to quantize the (generalized) cylinder $\bR^4 \times \bT^6$. 

Another motivation to quantize the cylinder comes from an idea of Kamani. In \dcite{11}, he studied the worldsheet of a superstring in a background as described before, as a noncommutative geometry. In this case, one quantizes the worldsheet, which is an ordinary cylinder $\bR \times \bT$. 

Apart from such physical arguments, the quantization of the cylinder is also interesting from a mathematical point of view. It turns out that the $C^\ast$-algebras occurring in the quantization of the plane and of the torus are rather different. As the cylinder in some sense lies in between the plane and the torus, it will be interesting to study the $C^\ast$-algebras occurring in its quantization. Furthermore, the noncommutative cylinder  provides another example in the scarce list of finite-dimensional noncommutative geometries \cite{103}.

Having obtained the deformation quantization of the cylinder, it is interesting to consider its K-theory. This requires the K-theory of $C^\ast$-algebras, which turns out to be the right noncommutative analogue of topological K-theory. In fact, for the $C^\ast$-algebra $C(X)$ of continuous functions on a compact Hausdorff space $X$ we have that
\bd
K_n \big (C(X) \big) = K^n (X).
\ed
The main results in topological K-theory, like Bott periodicity and homotopy invariance, lift to the K-theory of $C^\ast$-algebras which has the additional powerful feature of stability \cite{rordam}. 

It is interesting to study the interplay between K-theory and deformation quantization. We say that K-theory is \textbf{rigid} under a given deformation when $K(\cA_\hbar)$ is independent of $\hbar$ \cite{104}. For example, both Bott periodicity \cite{105} and a far-reaching generalization of it, the Baum-Connes conjecture in E-theory \cite{connes} can be seen as examples of such rigidity \cite{91}. For the three examples just mentioned, {i.e.}, Euclidean space, the torus and the cylinder, rigidity of K-theory turns out to hold. However, in general this is not the case. Let $\cA_0= C_0(T^\ast M)$ and $\cA_\hbar = \cB_0(L^2(M) )$ for al $\hbar >0$. Then for general $M$, clearly 
\bd
K_n(\cA_0) = K^n (T^\ast M) \neq K_n (\cA_\hbar) = \left\{ \begin{array}{ll} \bZ & \mbox{if } n=0 \\ 0 & \mbox{if } n=1  \end{array} \right.
\ed

The Gel'fand-Naimark theorem assures us that we can obtain all topological notions of a locally compact Hausdorff space from the $C^\ast$-algebra of continuous functions on it. However, in order to describe the full geometry of a spin manifold, we need more data. It turns out that the right algebraic description of a spin manifold is given by a real spectral triple satisfying Connes' seven axioms.
\begin{defn}
A \textbf{spectral triple} $\triple$ is given by a unital involutive algebra of operators $\cA$ on a Hilbert space $\cH$ and a self-adjoint operator $D=D^\ast$ on $\cH$ such that
\begin{enumerate}
\item The resolvent $(D-\lambda)^{-1}$ is compact for all $\lambda \notin \Sp(D)$,
\item The commutators $[D,a]:= Da-aD$ are bounded for any $a \in \cA$.
\end{enumerate}
\end{defn}
For the formulation of the seven axioms that define a \textbf{spin geometry} on $\cA$, we refer to \dcite{52,varilly,93}. A complete reconstruction of the spin structure on a spin manifold $M$ from the spin geometry over $C^\infty(M)$, can be found in \dcite{107}.

Spectral triples provide a powerful tool in describing noncommutative geometries but, at least in this definition, it relies heavily on two conditions, namely: 
\begin{itemize}
\item $1 \in \cA$,
\item $D$ is self-adjoint.
\end{itemize}
In (commutative) spin geometry, this is equivalent to the condition that $M$ is a compact Riemannian spin manifold \cite{107}. In physics, however, it is natural to work in a setting where this is not the case. This is illustrated by simple examples. Consider a Minkowski space-time $M=\bR^4$ with an indefinite metric $\eta=(-1, 1, 1, 1)$. The Dirac operator on $M$ is neither self-adjoint nor elliptic, and $M$ is noncompact. Other examples come from string theory. Consider the worldsheet of a string $\bR \times \bT$, embedded in a compactified background $\bR^4 \times \bT^6$. Both the worldsheet and the background have a semi-Riemannian metric, so that both conditions are unfulfilled. 

Thus, in order to describe such physical models in noncommutative geometry, {i.e.} using a spectral triple, we need to adjust its definition. If the $C^\ast$-algebra is nonunital, it is sufficient to replace condition 1 in the definition of a spectral triple by
\begin{itemize}
\item[1'.] The operator $a (D- \lambda)^{-1}$ is compact for any $a \in \cA$; $\lambda \notin \Sp(D)$,
\end{itemize}
However, in Lorentzian geometry, the Dirac operator $D$ is not self-adjoint, so that this condition must be dropped. It turns out that the operator $D$ is a Krein-self-adjoint operator in a Krein space $\cH$. 

Noncompact Lorentzian manifolds are central objects in physics and, therefore, we will discuss here the adjustifications mentioned in the definition of spectral triples. It will turn out that this can be done in a natural way, which allows for a definition of isospectral deformation, similar to what has been done by Connes and Landi \cite{92}. Our key example of a noncommutative noncompact Lorentzian manifold will be the noncommutative cylinder, which is defined by isospectral deformation of the Lorentzian cylinder. 
\\[5mm]
In section 2, we discuss deformation quantization of Euclidean space, the torus and the cylinder. We obtain the family of $C^\ast$-algebras as a family of crossed product algebras and discuss their K-theory. We provide a new evidence for the idea that K-theory is rigid under deformation quantization by describing the K-theory of the noncommutative cylinder. 

In section 3, still working in the Euclidean setting, we consider Connes' trace theorem for noncompact manifolds. We construct spectral triples for algebras without a unit and discuss Connes' character formula in the case of the cylinder. It turns out that it is possible to generalize this theorem to noncompact manifolds. Then we obtain the noncommutative cylinder as a spectral triple, via isospectral deformation of the canonical triple of the cylinder, similar to what is done by Connes and Landi \cite{92}. We attempt to construct a spin geometry over the noncommutative cylinder, where Connes' seven axioms are adapted to nonunital algebras as in \dcite{109}. 

We adjust the definition of the spectral triple to semi-Riemannian spin geometry \cite{99} in section 4, in particular to Lorentzian spin geometry. This involves Krein spaces, and we give a short introduction to the theory of these spaces and operators acting in them. Since the Dirac operator is not self-adjoint, we work with the associated operator $\Delta_J$, which is self-adjoint. It plays a central role in the formulation of the integral in terms of a Dixmier trace. We discuss two Hochschild cocycles that can be associated to the semi-Riemannian canonical triple. 

Finally, we consider the noncommutative Lorentzian cylinder, obtained by isospectral deformation of the semi-Riemannian spectral triple that describes the Lorentzian cylinder. The set of admissible fundamental symmetries for the noncommutative cylinder is shown to be exactly the set of fundamental symmetries coming from spacelike reflections in spinor space. 

\section{Deformation quantization and K-theory} 
\subsection{Old examples}
We start with a brief recapitulation of the definition of strict deformation quantization. Subsequently, we review the strict deformation quantization of Euclidean space and of the torus, both due to Rieffel \cite{80,90,83,86}. 
\begin{defn}
Let $M$ be a Poisson manifold with bracket $\{ \mbox{ } , \mbox{ } \}$ and let $\cA$ be a dense $\ast$-subalgebra of $C_0(M)$. A \textbf{strict deformation quantization} of $M$ in the direction of $\{ \mbox{ } \mbox{ } \}$, consists of an open interval $I \subseteq \bR$ with $0$ as an accumulation point, together with, for each $\hbar \in I$, an associative product $\ast_\hbar$, an involution $^{\ast_\hbar}$, and a $C^\ast$-norm $\| \mbox{ } \|_\hbar$ (for $\ast_\hbar$ and $^{\ast_\hbar}$) on $\cA$, which for $\hbar=0$ are the original pointwise product, complex conjugation involution, and supremum norm, such that
\begin{enumerate}
\item The family $\{ \cA_\hbar \}_{\hbar \in I}$ forms a continuous field of $C^\ast$-algebras over $I$. Here $\cA_\hbar$ denotes the $C^\ast$-completion of $\cA$ with respect to $\| \mbox{ } \|_\hbar$.
\item For every $f,g \in \cA$, 
\begin{displaymath}
\lim_{\hbar \to 0} \| (f \ast_\hbar g - g \ast_\hbar f)/{\ii \hbar} - \{ f, g\} \|_\hbar  =0. \quad \mbox{ (Dirac's condition)}
\end{displaymath}
\end{enumerate}
\end{defn}
\subsubsection{Weyl quantization}
We consider even-dimensional Euclidean space $\bR^{2n}$. Let $\cS(\bR^{2n})$ denote the commutative algebra of Schwartz functions on $\bR^{2n}$ under pointwise multiplication. This pointwise product is deformed to the Moyal star product, which reads, in Fourier space, for any $\hbar \in \bR$
\begin{equation} \label{starR}
(\phi \ast_\hbar \psi )(p,q) = \int_{\bR^{2n}} \dd^n p' \mbox{ } \dd^n q' \mbox{ } \phi(p',q') \psi(p-p', q-q') e^{-\ii \hbar( q' \cdot p - p'\cdot  q )}.
\end{equation}
The involution we use on $\cS(\bR^{2n})$ is defined by $\phi^\ast(p,q) = \overline{\phi(-p,-q)}$, which is independent of $\hbar$. We let $\pi_\hbar$ denote the left regular representation of $\cS(\bR^{2n})$ on $L^2(\bR^{2n})$ via $\ast_\hbar$, {i.e.}  for $\phi \in \cS(\bR^{2n})$ and $\Psi \in L^2(\bR^{2n})$, 
\begin{displaymath}
\pi_\hbar(\phi) \Psi :=  \phi \ast_\hbar \Psi.
\end{displaymath} 
We define a norm $\|\mbox{ } \|_\hbar$ on $\cS(\bR^{2n})$ as the operator norm for this representation. The completion of $\cS(\bR^{2n})$ with respect to this norm is a $C^\ast$-algebra, denoted by $\bR^{2n}_\hbar$. 
By rewriting formula (\ref{starR}) in terms of partial Fourier transforms, one can show the following \cite{landsman}. 
\begin{prop} \label{planecross}
The $C^\ast$-algebra $\bR^{2n}_\hbar$ is isomorphic to the crossed product algebra 
\begin{displaymath}
C_0(\bR^n) \rtimes_\hbar \bR^n, 
\end{displaymath}
where $\bR^n$ acts on $\bR^n$ by translation, $x \mapsto x+\hbar y$ ($x,y \in \bR^n$), so that it acts on $C_0(\bR^n)$ by the pullback of this action. $\Box$
\end{prop}
\begin{thm}
For $\hbar \neq 0$, the $C^\ast$-algebra $\bR^{2n}_\hbar$ is isomorphic to $\cB_0 (L^2(\bR^n))$, the $C^\ast$-algebra of compact operators on $L^2(\bR^n)$. 
\end{thm}
For a proof of this, we refer to \dcite{landsman,86}. 

It is now immediate that the $C^\ast$-algebras $\bR^{2n}_\hbar$ ($\hbar \neq 0$) are simple, and are all isomorphic to each other. Furthermore, we can conclude that Euclidean space $\bR^{2n}$ has rigid K-theory under quantization, {i.e.}, for all $\hbar$ one has
\begin{eqnarray*}
&K^0( \bR^{2n}) \cong K_0 ( \bR^{2n}_\hbar ) \cong \bZ;&\\
&K^1( \bR^{2n}) \cong K_1 ( \bR^{2n}_\hbar ) \cong 0.&
\end{eqnarray*}
\subsubsection{Noncommutative tori}
Let $\bT^d$ be the $d$-dimensional torus, and let $\theta$ be a real skew-symmetric $d \times d$ matrix. Instead of deforming the pointwise product in the space of smooth functions on $\bT^d$, we deform the product in its Fourier space $\cS(\bZ^d)$. 
For $\hbar \in \bR$, the star product reads
\begin{equation}
(\phi \ast_\hbar \psi ) (n) = \sum_{m \in \bZ^d} \phi(m) \psi(n-m) e^{2\pi \ii \hbar \theta (m,n) }.
\end{equation}
Here $\theta$ is the skew bilinear form defined by $\theta (m,n) := \sum_{j,k} \theta_{jk} m_j n_k$.

We set $\phi^\ast(n) := \overline{\phi(-n)}$, which is independent of $\hbar$. We let $\cS(\bZ^d)$ act on $L^2 (\bZ^d)$ by left multiplication via $\ast_\hbar$. The completion of $\cS(\bZ^d)$ with respect to the operator norm $\| \mbox{ } \|_\hbar$, equipped with this star product is the \textbf{noncommutative torus}, denoted by $\bT^d_{\hbar\theta}$. For fixed $\theta$, the family $\{ \bT^d_{\hbar\theta} \}$ provides a strict deformation quantization of $\bT^d$ \cite{80}. When $d=2$, the skew-symmetric matrix $\theta$ is just determined by a real number, denoted by $\theta$ as well. It turns out that the noncommutative torus $\bT^2_{\theta}$ is isomorphic to the crossed product algebra $C(\bT) \rtimes_\alpha \bZ$, where $\alpha (f)(t) := f(t+\theta)$. Furthermore, it is simple if and only if $\theta$ is irrational. If $\theta \neq \theta'$, both irrational with $0<\theta, \theta' <\frac{1}{2}$, then $\bT^2_\theta \ncong \bT^2_{\theta'}$\cite{varilly}. It came as a surprise that the K-groups of $\bT^d_\theta$ do not depend on $\theta$. 
\begin{prop} \label{torusrigid}
The torus $\bT^d$ has rigid K-theory under quantization, {i.e.}, for all $\hbar$ one has
\begin{eqnarray*}
&K^0( \bT^d) \cong K_0 ( \bT^d_{\hbar\theta} ) \cong \bZ^{2^{d-1}};&\\
&K^1( \bT^d) \cong K_1 ( \bT^d_{\hbar\theta} ) \cong \bZ^{2^{d-1}}.&
\end{eqnarray*}
\end{prop}
\subsection{Deformation quantization of cylinders}  \label{defnNCC}
We consider the cylinder in a generalized form. The 
\textbf{$(n,d)$-dimensional cylinder $C^{(n,d)}$} is defined as 
\begin{equation} 
C^{(n,d)}:= \bR^n \times \bT^d.
\end{equation} 
In the case $n=d=1$ we obtain $C^2:=\bR \times S^1$, which is of course the familiar two-dimensional cylinder. 

Let $\Lambda$ be a Poisson structure on $\bR^n \times \bT^d$. For $j=1,\cdots, n+d$, let 
$\partial_{x_j}$ denote the vector field on $\bR^n \times \bT^d$ corresponding to 
differentiation in the $j^{\mathrm{th}}$ direction. We can write the Poisson structure 
as 
\begin{equation} 
\Lambda=-\pi^{-1} \sum_{i<j} \theta_{ij} \partial_{x_i} \wedge \partial_{x_j}. 
\end{equation} 
The factor $\pi^{-1}$ has been included for later convenience. Here $\theta_{ij}$ is a real 
skew-symmetric matrix. For later use, we define a skew bilinear form $\theta$ on $\bR^n \times \bZ^d$, 
\begin{equation}\label{biform} 
\theta(l,k)= \sum_{i,j} \theta_{ij} l_i k_j, \qquad (l,k \in \bR^n \times \bZ^d). 
\end{equation} 
Let $\lambda$ denote Lebesgue measure on $\bR^n \times \bT^d$. The Fourier transform 
$\hat{f}$ of a Schwartz function $f \in \cS(\bR^n \times \bT^d)$ is given by 
\begin{equation}\label{NCC:Fourier} 
\hat{f} (k) = \int_{\bR^n \times \bT^d} \dd \lambda (x) e^{-2 \pi \ii k \cdot x} f(x). 
\end{equation} 
For $i=1,\ldots, n$ we have $k_i \in \bR$, for $i=n+1, \ldots, n+d$ we have $k_i \in 
\bZ$. In fact, the Fourier transform maps $\cS(\bR^n \times \bT^d)$ isomorphically to $\cS(\bR^n \times \bZ^d)$.

To integrate over $\bR^n$ and sum over $\bZ^d$ in the product $\bR^n \times \bZ^d$, we introduce the measure $\mu$ on $\bR^n \times \bZ^d$, defined as the product of Lebesgue measure on $\bR^n$ and the counting measure on $\bZ^d$. 

For functions in Fourier space, the Poisson bracket is given by 
\begin{eqnarray}\label{NCC:Poissonbracket} 
\{ \phi, \psi \}(k) &=& 4\pi \int_{\bR^n \times \bZ^d} \dd \mu(l) \sum_{i,j} 
\theta_{ij} l_i \phi(l) (k_j-l_j) \psi(k-l) \\ \nonumber &=& 4\pi \int_{\bR^n \times 
\bZ^d} \dd \mu(l)\mbox{ } \phi(l) \psi(k-l) \theta(l,k) 
\end{eqnarray} 
where $k,l \in \bR^n\times \bZ^d$ and $\theta$ is the bilinear form  defined in 
equation (\ref{biform}). 

We define a bicharacter $\sigma_\hbar$ on $\bR^n\times \bZ^d$ by 
\begin{equation} 
\sigma_\hbar (l,k) = e^{2\pi \ii \hbar \theta(l,k)}, 
\end{equation} 
where $\hbar \in \bR$, and introduce a star product $\ast_\hbar$ on $\cS(\bR^n\times \bZ^d)$ by
\begin{equation} \label{NCC:starproduct} 
(\phi \ast_\hbar \psi) (k) = \int_{\bR^n \times \bZ^d} \dd \mu(l) \mbox{ } \phi(l) 
\psi(k-l) \sigma_\hbar(l,k).
\end{equation} 
We define an involution on $\cS(\bR^n \times \bZ^d)$ by $\phi^\ast (k) := \overline{\phi(-k)}$, independent of $\hbar$. We represent $\cS(\bR^n \times \bZ^d)$ on $L^2(\bR^n \times \bZ^d)$ by star product multiplication, and define the \textbf{noncommutative cylinder} as the completion of $\cS(\bR^n \times \bZ^d)$ in the operator norm $\| \mbox{ } \|_\hbar$, equipped with product $\ast_\hbar$. This $C^\ast$-algebra is denoted by $C^{(n,d)}_{\hbar\theta}$.

We could equally well define the noncommutative cylinder as the (completion of) the 
algebra $\cS(\bR^n \times \bT^d)$ with product, involution and norm obtained by pulling 
back the product $\ast_\hbar$, involution $^\ast$ and norm $\| \mbox{ } \|_\hbar$ through the inverse Fourier transform. Even though this makes the differences with the ordinary cylinder more clear, we will continue in Fourier space to avoid expressions involving many derivatives. 
\begin{thm}
For fixed $\theta$, the family $\big\{ C^{(n,d)}_{\hbar\theta} \big\}$ provides a strict deformation quantization of $\bR^n \times \bT^d$ in the direction of $\{ \mbox{ } , \mbox{ } \}$. 
\end{thm}
\textit{Proof. } 
We verify Dirac's condition 
\begin{equation} 
\| (\phi \ast_\hbar \psi - \psi \ast_\hbar \phi)/{\ii \hbar} - \{ \phi, \psi \} 
\|_\hbar \to 0 \quad \mbox{ as } \hbar \to 0, 
\end{equation} 
where $\phi,\psi \in \cS(\bR^n \times \bZ^d)$. We define 
\begin{displaymath} 
\Delta_\hbar :=(\phi \ast_\hbar \psi - \psi \ast_\hbar \phi)/{\ii \hbar} - \{ \phi, 
\psi \}. 
\end{displaymath} 
With formulae (\ref{NCC:Poissonbracket}) and (\ref{NCC:starproduct}), this reads 
\begin{displaymath} 
\Delta_\hbar (k)  =\int_{\bR^n \times \bZ^d} \dd \mu(l) \mbox{ } \phi(l) \psi(k-l) 
\big( (\sigma_\hbar(l,k)-\sigma_\hbar(k,l))/{\ii \hbar} - 4\pi \theta(l,k) \big). 
\end{displaymath} 
Similar to Rieffel in \dcite{80}, we can estimate the expression inside $\big( \mbox{ } 
\big)$ so that, 
\begin{displaymath}    
| \Delta_\hbar (k) | \leq \hbar M \int_{\bR^n \times \bZ^d} \dd \mu(l) \mbox{ } 
|\phi(l)| \mbox{ } |\psi(k-l)| |l|^2 |k-l|^2,
\end{displaymath} 
for some constant $M$. This last expression is just (proportional to) the convolution 
product of two functions $\widetilde{\phi}$ and $\widetilde{\psi}$ where 
\begin{displaymath} 
\widetilde{\phi}(k):= |k|^2 |\phi(k)|, \quad \widetilde{\psi}(k):= |k|^2 |\psi(k)|. 
\end{displaymath} 
As the $L^1$-norm dominates the norm $\| \mbox{ } \|_\hbar$, we have 
\begin{displaymath} 
\| \Delta_\hbar \|_\hbar \leq \hbar M \| \widetilde{\phi} \ast \widetilde{\psi} \|_1.
\end{displaymath} 
It follows that $\| \Delta_\hbar \|_\hbar \to 0$ as $\hbar \to 0$. 
\\
Continuity of the field $\{ C^{(n,d)}_{\hbar\theta} \}$ follows from Corollary 5.6 in \dcite{106} (or Lemma 1 in \dcite{90}), in combination with Proposition \ref{cylcross} below. $\Box$

\subsection{Properties of noncommutative cylinders} \label{PropNCC}
When one observes the major differences between $\bR^{2n}_\hbar$ and $\bT^d_{\hbar\theta}$, one is led to the questions whether the noncommutative cylinders are simple and whether they are all isomorphic. For this, we connect with the theory of crossed product algebras. At the end, we discuss the K-theory of noncommutative cylinders. 

We take the noncommutative cylinder for $n=d$, and denote it by $C_\hbar^{2d}$. We let $l=(x, n)$ and $k=(y, m)$, where $x,y \in \bR^d$ and $n,m \in \bZ^d$, and choose the following skew bilinear form on $\bR^d \times \bZ^d$,
\begin{equation}
\theta(l,k) =\frac{1}{2\pi} \sum_{i=1}^d y_i n_i - m_i x_i.
\end{equation}
We want to rewrite the star product (\ref{NCC:starproduct}) in terms of partial Fourier transforms, defined by
\begin{equation}
\acute{\phi} (x,t) := \sum_{n \in \bZ^d} \phi(x,n) e^{\ii n \cdot t} \qquad (t\in \bT^d),
\end{equation}
which is a function on $\bR^d \times \bT^d$. The star product on $\cS(\bR^d \times \bT^d)$ then reads
\begin{equation}
(\acute{\phi} \ast_\hbar \acute{\psi} )(x,t) = \int_{\bR^d} \dd y \mbox{ }\acute{\phi} (y, t+ \hbar(y-x) ) \mbox{ } \acute{\psi}(x-y, t+\hbar y),
\end{equation}
as can be easily verified. We introduce an action $\beta$ of $\bR^d$ on $\bT^d$ defined by $\beta_x (t) = t+ \hbar x$, and write
\begin{equation}
(\acute{\phi} \ast_\hbar \acute{\psi} )(x,t) = \int_{\bR^d} \dd y \mbox{ } \acute{\phi} (y, \beta_{y-x}(t) ) \mbox{ } \acute{\psi}(x-y, \beta_y(t)).
\end{equation}
This formulation of the star product in terms of an action $\beta$ of $\bR^d$ on $\bT^d$ goes back to Rieffel. As in the examples in his paper \cite{86}, we relate this to crossed product algebras. For more details on the theory of these algebras, we refer to Pedersen \cite{pedersen2}. Let $C(\bT^d) \rtimes_\hbar \bR^d$ denote the crossed product algebra for the $\hbar$-dependent action $\beta_{2x}$. Then $\cS(\bR^d,C^\infty(\bT^d) )$ is a dense $\ast$-subalgebra of this crossed product algebra. Define a map $Q: \cS(\bR^d \times \bT^d) \to \cS(\bR^d,C^\infty(\bT^d) )$ by
\begin{equation}
Q(\acute{\phi})(x,t) := \acute{\phi} (x,\beta_x(t)). 
\end{equation}
Note that $\cS(\bR^d \times \bT^d)$ is a dense $\ast$-subalgebra of $C_\hbar^{2d}$. Clearly, $Q$ is an isomorphism, in that
\begin{eqnarray}
Q(\acute{\phi} \ast_\hbar \acute{\psi} )(x,t) &=& (\acute{\phi} \ast_\hbar \acute{\psi})(x, \beta_x(t)) \\ \nonumber
&=& \int_{\bR^d} \dd y \mbox { } Q(\acute{\phi})(y,t) \mbox{ } Q(\acute{\psi}) (x-y,\beta_{2y}(t) )\\ \nonumber
&=& Q(\acute{\phi}) \ast Q(\acute{\psi}).
\end{eqnarray}
Extension of the map $Q$ to $C_\hbar^{2d}$ yields the following. 
\begin{prop} \label{cylcross}
The noncommutative cylinder $C_\hbar^{2d}$ ($\hbar \neq 0$) is isomorphic to the crossed product $C(\bT^d) \rtimes_\hbar \bR^d$. $\Box$
\end{prop}
This allows us to use known results on crossed product algebras.
\begin{thm}
The $C^\ast$-algebra $C^{2d}_\hbar$ is isomorphic to $\cB_0 (L^2(\bT^d)) \otimes C^\ast(\bZ^d)$. 
\end{thm}
\textit{Proof. }We note that $C(\bT^d) \rtimes_\hbar \bR^d \cong C(\bT^d) \rtimes_{\hbar'} \bR^d$ for $\hbar, \hbar' \neq 0$. In particular, 
\begin{displaymath}
C(\bT^d) \rtimes_\hbar \bR^d \cong C(\bT^d) \rtimes \bR^d
\end{displaymath}
for $\hbar \neq 0$. Corollary $2.8$ of Green \cite{87} completes the proof. $\Box$
\\[5mm]
With the isomorphism $C^\ast(\bZ^d) \cong C(\bT^d)$, we have the following. 
\begin{corl}
The noncommutative cylinders $C^{2d}_\hbar$ ($\hbar \neq 0$) are nonsimple $C^\ast$-algebras. $\Box$
\end{corl}
It is well known that any $C^\ast$-algebra $A$ is Morita equivalent to its stabilization
\begin{displaymath}
A_S:=\cB_0 (\cH)  \otimes A
\end{displaymath}
for some Hilbert space $\cH$. In particular, $\cB_0 (L^2(\bT^d)) \otimes C(\bT^d)$ is Morita equivalent to $C(\bT^d)$. Since Morita-equivalent $C^\ast$-algebras have isomorphic K-groups, we have the following.
\begin{corl}
For the noncommutative cylinder $C_\hbar^{2d}$ one has for all $\hbar \neq 0$,
\begin{displaymath}
K_0(C_\hbar^{2d}) \cong K_1(C_\hbar^{2d})\cong \bZ^{2^{d-1}}.
\end{displaymath}
\end{corl}
In order to compare this with the K-groups of the original cylinder $\bR^d \times \bT^d $ we need the following Lemma. 
\begin{lma} \label{cylKgroups}
For the cylinder $\bR^d \times \bT^d $ the K-groups are
\begin{displaymath}
K^0 ( \bR^d \times \bT^d ) \cong K^1 (\bR^d \times \bT^d ) \cong \bZ^{2^{d-1}}.
\end{displaymath}
\end{lma}
\textit{Proof. }
For the K-groups of $\bT^d$ we note that
\begin{displaymath}
K_0(C(\bT, \cA) ) \cong K_1(C(\bT, \cA) ) \cong K_0(\cA) \oplus K_1(\cA)
\end{displaymath}
for any $C^\ast$-algebra $\cA$ (cf. Exercise 10.1 in \dcite{rordam}). Since 
\bd
C_0 (\bR^d \times \bT^d) \cong C(\bT, C_0 (\bR^d \times \bT^{d-1}) ),
\ed
this yields by induction
\begin{displaymath}
K_0(C_0(\bR^d \times \bT^d) ) \cong K_1(C_0(\bR^d \times \bT^d)) \cong \bZ^{2^{d-1}}.
\end{displaymath}
Here, one uses $K_0(C_0(\bR^d)) \oplus K_1(C_0(\bR^d)) \cong \bZ$. 
An alternative proof can be constructed using Bott periodicity. $\Box$
\begin{prop}
The cylinder $\bR^d \times \bT^d$ has rigid K-theory under quantization, {i.e.}, for all $\hbar$ one has
\begin{eqnarray*}
&K^0( \bR^d \times \bT^d) \cong K_0 ( C^{2d}_\hbar ) \cong \bZ^{2^{d-1}};&\\
&K^1( \bR^d \times \bT^d) \cong K_1 ( C^{2d}_\hbar ) \cong \bZ^{2^{d-1}}.&
\end{eqnarray*}
\end{prop}
\begin{flushright}
$\Box$
\end{flushright}
Note that these groups are the same as the K-groups of the noncommutative torus $\bT^d_{\hbar\theta}$.
\section{Noncommutative manifolds and isospectral deformation}
The description of a manifold in terms of spectral data is provided by the theory of $K$-cycles (also called spectral triples) developed by Connes. This generalization has been very successful in describing noncommutative manifolds, as shown by examples like the noncommutative torus \cite{connes} and the noncommutative $4$-sphere $S^4_\theta$ \cite{92}. It also admits generalizations to noncompact manifolds, or, in other words, to nonunital algebras. 
\subsection{Connes' trace theorem}
An important result here is that Connes' trace theorem generalizes to noncompact manifolds\cite{varilly}. Connes' trace theorem\cite{connes,95} relates the Wodzicki residue of an elliptic pseudodifferential operator to the Dixmier trace of this operator. It allows one to compute the integral of any function on a compact Riemannian manifold in terms of an operatorial formula. See for example \dcite{varilly} for a complete treatment and proof of the theorem. 
\begin{prop} \label{connestrace}
Let $f$ be an integrable function on an $n$-dimensional Riemannian manifold $M$, then
\begin{displaymath}
\int_M f(x) \sqrt{| g|} \dd x = \frac{n (2 \pi)^n}{\Omega_n} \trace_\omega (f \Delta^{-n/2}).
\ed
where $\Delta$ is the Laplacian on $M$. 
\end{prop}
The fact that a manifold $M$ is not compact translates into the fact that the $C^\ast$-algebra $C_0(M)$ is not unital. So, in order to describe a Riemannian manifold which is only locally compact by a spectral triple, we need a generalization of the definition as given in \dcite{connes}. 
\begin{defn} \label{TripleNonUn}
A spectral triple $\triple$ is given by an involutive algebra of operators $\cA$ in a Hilbert space $\cH$ and a self-adjoint operator $D=D^\ast$ in $\cH$ such that
\begin{enumerate}
\item $a (D-\lambda)^{-1}$ is compact for any $a \in \cA$; $\lambda \notin sp(D)$,
\item The commutators $[D, a] := Da-aD$ are bounded for any $a \in \cA$.
\end{enumerate}
The triple is said to be \textbf{even} if there is a $\bZ_2$ grading of $\cH$, namely an operator $\chi$ on $\cH$ with $\chi^\ast= \pm \chi$ and $\chi^2=1$, such that
\bea
\chi D + D \chi &=& 0, \nonumber \\
\chi a - a \chi &=& 0  , \qquad \mbox{for all } a \in \cA.
\eea
If such a grading does not exist, the triple is said to be \textbf{odd}, and we set $\chi=1$. 
\end{defn}

This was already pointed out by Connes in 1995 (\dcite{97}). If $\cA$ is unital this yields the familiar definition, because then condition 1 implies that $1_\cA (D-\lambda)^{-1}$ is compact.

As a special case, we have the Dirac geometry $(C^\infty_0(M), \mbox{ }L^2 (M,S),\mbox{ } \Dirac)$ where $M$ is a spin manifold and $\Dirac$ the Dirac operator for a spin bundle $S \to M$. Here, $C^\infty_0(M)$ denotes the algebra of smooth continuous functions on $M$ 'vanishing at infinity'. For general $M$ this means for $f \in C_0(M)$ that for all $\epsilon >0$, there exists a compact submanifold $K$ of $ M$ such that $f(x) <\epsilon$ for all $x \in M/K$. 
Note that the principal symbol $\sigma(\Delta)$ of the Laplacian, coincides with $\sigma(\Dirac^2)$. 

\begin{defn}
A spectral triple $\triple$ is said to be \textbf{$p^+$-summable} ($p>0$), if $a |D|^{-p} \in \cL^{(1,\infty)}$ for any $a \in A$ for some dense subalgebra $A \subset \cA$.
\end{defn}
For a $p^+$-summable spectral triple $\triple$, the \textbf{noncommutative integral} of $a \in \cA$ is defined by
\be
\intC a := \frac{n(2\pi)^n}{2^{[n/2]}\Omega_n} \trace_\omega a |D|^{-n}.
\ee 
If $\cA = C^\infty_0(M)$, the $\ast$-subalgebra $A$ consists of integrable functions with respect to the measure associated to the Riemannian metric on M.

\subsection{Connes' character formula for the cylinder} \label{conneschar}
Another result in noncommutative geometry is Connes' character formula. It provides a link between Hochschild and cyclic cohomology in that it gives a representation of the Hochschild class of the Chern character, {i.e.}, a cyclic cocycle. It turns out that the Hochschild cocycle is much easier to handle than the Chern character. Here, we prove the character formula for the cylinder. As we saw in the previous section, the geometry of the $(n,d)$-dimensional cylinder can be described by the triple
\begin{eqnarray*}
\cA &:=& C^\infty_c(\bR^n \times \bT^d  )\\
\cH &:=& L^2 (\bR^n \times \bT^d ) \otimes \bC^{2^{[(n+d)/2]} } \\
D &:=& \Dirac
\end{eqnarray*}
For convenience we have restricted $\cA$ to functions with compact support, so that all functions in $\cA$ are integrable.

The Dirac operator on the cylinder is defined by $\Dirac := \gamma^a \partial_a$ where the gamma-matrices satisfy $\{ \gamma^a, \gamma^b \} = 2\delta^{ab}$. Using the spectral theorem for self-adjoint operators we define $F = \sign (\Dirac)$, where $\sign(x) = +1(-1)$ for $x \geq 0$ ($x<0$). The couple $(\cH, F)$, together with a representation $\sigma$ of $\cA$ in $\cH$, defines a Fredholm module over $\cA$. In the case $n+d$ is even, there is grading operator on $\cH$ defined by $\chi := \ii^{n+d} \gamma^1 \cdots \gamma^{n+d}$, which makes $(\cH, F)$ an even Fredholm module. Before we continue, we state some theory on universal differential graded algebras for nonunital algebras, which will be needed later. 
\subsubsection{Universal forms on nonunital algebras}
The way to describe the graded differential algebra for nonunital algebras is very similar to the way K-theory is defined for nonunital algebras. Both rely on the notion of unitization. For the theory of universal graded differential algebras for algebras with unit, we refer to  \dcite{varilly,15,94}. A comprehensive introduction is found in Chapter 4 of \dcite{22}. The approach we take here is based on \dcite{94} and \dcite{98}.

Let $\cA$ be an algebra. Its unitization $\widetilde{\cA}$ is defined by $\widetilde{\cA}:= \cA \oplus \bC$. The quotient map is $\pi : \widetilde{\cA} \to \bC$ with $\cA= \ker(\pi)$. Since $1 \in \widetilde{\cA}$, we can construct the graded universal differential algebra $\Omega \widetilde{\cA}$ following standard literature. The relation between the differential algebras is similar to K-theory, {i.e.}, $\Omega \widetilde{\cA} = \bC \oplus \Omega {\cA}$. Let $\delta$ be the corresponding derivation. By Proposition 3.2 of \dcite{94} we can extend $\pi$ uniquely to a map $\pi_\ast : \Omega \widetilde{\cA} \to \Omega \bC$ by
\be
\pi_\ast ( \tilde{a}_0 \mbox{ } \delta \tilde{a}_1 \cdots \delta \tilde{a}_n ) = \pi \tilde{a}_0 \mbox{ } \delta(\pi \tilde{a}_1) \cdots \delta(\pi \tilde{a}_n)
\ee
Immediately, this yields $\pi_\ast (\Omega^n\widetilde{\cA}) = \{ 0 \}$ for $n>0$. For $n=0$, we have $\pi_\ast(a + \lambda 1) =\lambda $. Of course, $\Omega \bC = \bC$.  

Similar to what is done in the definition of K-theory for nonunital algebras, we define the graded universal differential algebra $\Omega \cA$ of the nonunital algebra $\cA$ as a kernel
\be
\Omega \cA := \ker (\pi_\ast : \Omega \widetilde{\cA} \to \Omega \bC).
\ee
From the above observations it is clear that $\Omega^n \cA = \Omega^n \widetilde{\cA}$ for $n>0$. For $n=0$, we have $\Omega^0 \widetilde{\cA} = \bC \oplus \Omega^0{\cA}$, which yields $\Omega^0 \cA = \cA$. 
\subsubsection{The Chern character}
Given a Fredholm module $(\cH, F, \sigma)$ over $\cA$, we will construct a representation of $\Omega \cA$, for $\cA=C^\infty_c(\bR^n \times \bT^d  )$. This will use the fact that the map
\bd
\dd : a \mapsto \ii [ F, \sigma(a)]
\ed
is a derivation on $\cA$, which commutes with the convolution. We can uniquely extend $\sigma$ to a representation of the unitization $\widetilde{\cA}$. The couple $(\cH, F)$ is a Fredholm module over $\widetilde{\cA}$, since obviously $[F, \tilde{a}] = [F, a]$ 
where we have suppressed the representations $\sigma$ and $\tilde{\sigma}$. We extend $\tilde{\sigma}$ to the universal differential algebra $\Omega\cA$ by
\begin{eqnarray*}
\tilde{\sigma}: \qquad \qquad \Omega^k \cA &\to& \cB(\cH) \\
\tilde{a}_0 \mbox{ } \delta \tilde{a}_1 \cdots \delta \tilde{a}_k &\mapsto&  \tilde{a}_0  \mbox{ }  \ii^k [F,  \tilde{a}_1] \cdots [F,  \tilde{a}_k]
= \tilde{a}_0 \mbox{ }\dd a_1 \cdots \dd a_k.
\end{eqnarray*}
From \dcite{varilly} we take the following lemma, generalized to the nonunital case. 
\begin{lma}
Let $\Dirac$ be the (Euclidean) Dirac operator on the cylinder $\bR^n \times \bT^d $ and $F= \sign(\Dirac)$. Then $[F, \sigma(a)] \in \cL^q (\cH)$, where $q=n+d+1$. $\Box$
\end{lma}
Besides the Schatten ideals $\cL^p(\cH)$, we define the conditional trace class $\cL^1_C (\cH)$ by
\be
\cL^1_C (\cH) := \big\{ a \in \cB(\cH) : a+FaF \in  \cL^1 (\cH) \big\}.
\ee
For elements in $\cL^1_C (\cH)$, we define the conditional trace by
\be
\trace_C (a) := \frac{1}{2} \trace(a+FaF).
\ee
\begin{defn}
The Chern character of the Fredholm module $(\cH, F)$ is the cyclic $(n+d)$ cocycle
\bd
\tau_F^{(n+d)} (\tilde{a}_0, a_1, \ldots, a_{n+d}) = \trace_C (\tilde{a}_0 \dd a_1 \cdots \dd a_{n+d}  ) \qquad (a_i \in \cA)
\ed
(with $\chi \tilde{a}_0$ instead of $\tilde{a}_0$ if $n+d$ is even).
\end{defn}
Note that since $\dd a_i = \ii [F, a_i] \in \cL^q(\cH)$, the above expression is indeed in $\cL^1_C (\cH)$, using H\"olders inequality. The following lemma brings us back from $\widetilde{\cA}$ to $\cA$.
\begin{lma}
\bd
\tau_F^{(n+d)} (\tilde{a}_0, a_1, \ldots, a_{n+d}) = \tau_F^{(n+d)} (a_0, a_1, \ldots, a_{n+d})
\ed
\end{lma}
\textit{Proof. }
This follows immediately by writing out $a + FaF$ for $a = \tilde{a}_0 \dd a_1 \cdots \dd a_{n+d}$.
\rightbox

Thus, although the Chern character is defined as a cyclic $(n+d)$-cocycle on $\widetilde{\cA}$, it is essentially a cyclic $(n+d)$-cocycle on $\cA$. 
\begin{thm} \label{langmann}
For all $a_0, a_1, \ldots ,a_{n+d} \in C^\infty_c(\bR^n \times \bT^d)$, one has 
\bd
\tau_F^{(n+d)} (a_0, a_1, \ldots, a_{n+d}) = c_{n+d} \int_{\bR^n \times \bT^d } a_0 \mbox{ } \dd_{dR} a_1 \wedge \cdots \wedge \dd_{dR} a_{n+d}.
\ed
for some constant $c_{n+d}$ and where $\dd_{dR}$ is the de Rham differential. $\Box$
\end{thm}
The proof of this theorem is really analogous to the case of $\bR^n$, which was proved by Langmann in \dcite{96}. 
\subsubsection{Connes' character formula}
It was already mentioned in \dcite{varilly} that Connes' character formula holds for noncompact manifolds. Specifically, we can construct a Hochschild $(n+d)$-cocycle, which agrees with the Chern character $\tau_F^{(n+d)}$ on Hochschild $(n+d)$-cycles to obtain Connes' character formula for the cylinder. In this simple case, this follows directly from Proposition \ref{connestrace} and Theorem \ref{langmann}, if we define a Hochschild $(n+d)$-cocycle by
\bd
\psi_D^\omega (\tilde{a}_0, a_1, \ldots, a_{n+d} ) := \trace_\omega (\tilde{a}_0 [\Dirac, a_1] \cdots [\Dirac, a_{n+d}] |\Dirac|^{-n-d} )
\ed
(with $\chi \tilde{a}_0$ instead of $\tilde{a}_0$ in the even case). This expression makes sense by the summability properties of the spectral triple. Similar to the Chern character, we can replace $\tilde{a}_0$ by $a_0$. 
\begin{thm}
For all $a_0, a_1, \ldots, a_{n+d} \in C^\infty_c(\bR^n \times \bT^d)$, we have
\bd
\psi_D^\omega (a_0, a_1, \ldots, a_{n+d} ) = \tau_F^{(n+d)} (a_0, a_1, \ldots, a_{n+d})
\ed
\end{thm}
\subsection{Isospectral deformation of the cylinder} \label{IsoDefR}
In section \ref{defnNCC} and \ref{PropNCC}, we constructed the noncommutative cylinder as the family of $C^\ast$-algebras $C_{\hbar\theta}^{(n,d)}$, $\hbar \in (0,1]$. Here $C_{\hbar\theta}^{(n,d)}$ was defined as the completion of $\cS(\bR^n \times \bT^d)$ with respect to the norm $\| \mb \|_\hbar$, equipped with product $\ast_\hbar$. Note that we here choose the more natural Fourier transform instead of $\cS(\bR^n \times \bZ^d)$. We now want to describe the geometry of the noncommutative cylinder using the theory of spectral triples developed in the previous sections. The approach we take involves isospectral deformation (see below) of the (Euclidean) Dirac geometry of the cylinder $\bR^n \times \bT^d$. 

Deformation quantization provides a natural technique to obtain a noncommutative analogue of a function algebra. Starting from the Dirac geometry $(C^\infty(M), \cH, D)$ of a compact spin manifold $M$, the simplest noncommutative manifold is the triple $(\cA_\hbar, \cH, D)$, where $\cA_\hbar$ is obtained from $C^\infty(M)$ along the lines of deformation quantization. This recipe for noncommutative manifolds is called isospectral deformation, since $\cH$ and $D$ are unchanged. The only thing that changes is the algebra and the way it acts on the Hilbert space. 

The Dirac geometry of the cylinder is given by $( C^\infty_0(\bR^n \times \bT^d), \mb \cH, \mb \Dirac)$, where $\cH = L^2(\bR^n \times \bT^d) \otimes \bC^{2^{[(n+d)/2]} }$. However, in order to represent the deformed algebra on the same Hilbert space, we have to restrict to $\cS(\bR^n \times \bT^d) \subset C^\infty_0(\bR^n \times \bT^d)$. Let $\cC_{\hbar\theta}^{(n,d)}$ denote the algebra $\cS(\bR^n \times \bT^d)$ equipped with product $\ast_\hbar$. Note the difference with the noncommutative cylinder $C_{\hbar\theta}^{(n,d)}$ as a $C^\ast$-algebra which is the completion of $\cC_{\hbar\theta}^{(n,d)}$. 

We want to construct a spin geometry on the noncommutative cylinder. Recall that a noncommutative spin geometry is a real spectral triple fulfilling Connes' seven axioms\cite{52}. It was shown in \dcite{varilly} and \dcite{110}, that when the algebra $\cA = C^\infty(M)$ on a compact spin manifold $M$, the spin structure, metric and Dirac operator can be recovered from these seven axioms. However, in our case, we need a modification of these conditions. Clearly, both $\cS(\bR^n \times \bT^d)$ and $\cC_{\hbar\theta}^{(n,d)}$ are nonunital. Hence, the conditions for a spin geometry on the noncommutative cylinder have to be modified. Such a definition of a noncommutative noncompact spin geometry has been proposed in \dcite{109} and \dcite{115}. 

Let us start by completing the set of ingredients for a spin geometry on the noncommutative cylinder. The basic element is the spectral triple $(\cC_{\hbar\theta}^{(n,d)}, \cH, \Dirac)$. A lengthy computation, similar to the one in section 4.1 of \dcite{115}, shows that $\pi_\hbar(f) (D-\lambda)^{-1}$ is indeed compact. Both the grading operator $\chi$ and the charge conjugation operator $C$ are inherited from the commutative case. 

Regularity, finiteness and reality follow directly from the commutative case, since we are considering an isospectral deformation. However, the classical dimension cannot be obtained directly from the spectrum of $\Dirac$, since the latter is continuous. Following \dcite{109}, it can be extracted from the leading term of the distributional kernel of $|\Dirac|$ \cite{110}, so that it follows from the commutative case, i.e. the classical dimension is $n+d$. The first order axiom is fulfilled since $[\Dirac, f] \in \cS(\bR^n \times \bT^d)$ if $f \in \cS(\bR^n \times \bT^d)$ and 
\bd
C \pi_\hbar(f^\ast) C^{-1} \psi= \psi \ast_\hbar f = \pi_\hbar^\circ (f) \psi.
\ed

For the orientation, we need a Hochschild $(n+d)$-cycle $c$ that satisfies $\pi_{\Dirac}(c) = \chi$. However, $\cC_\hbar^{2d}$ is Morita equivalent to $C^\infty(\bT^d)$ (see Appendix A) so that with Loday \cite{loday}
\be
HH(\cC_\hbar^{2d}) = HH(C^\infty(\bT^d))= H_{dR} (\bT^d)
\ee 
Hence, in the case $n=d$, $\pi_{\Dirac}(c) = 0$ for any $2d$-cycle.

Poincar\'e duality is satisfied in the case $n=d$ and the special form of $\theta$ described before. Since we are considering an isospectral deformation of $\bR^d \times \bT^d$, Poincar\'e duality follows from the commutative case. Indeed, we have
\bd
K_\bullet(\cC^{2d}_\hbar) \cong K^\bullet (\bR^d \times \bT^d)
\ed
as proved before. Since the pairing in Poincar\'e duality involves only the K-groups of the algebra and the Dirac operator $\Dirac$, the claim follows. 

The real spectral triple $\big( \cC_\hbar^{2d}, \mb\cH, \mb\Dirac, C, \chi \big)$ satisfies almost all conditions for a spin geometry on $\cC_\hbar^{2d}$. Only the orientation class does not exist for the dimension prescribed by the Dirac operator. This illustrates again \cite{109,115} the need for an adjustment of the orientation axiom. 
\section{Noncommutative Lorentzian manifolds and isospectral deformation} \label{Lorentzcase}
We saw in the previous sections that spectral triples provide a powerful noncommutative description of Riemannian geometry, which allows for generalizations to noncompact manifolds. However, in physics it is more natural to consider Lorentzian manifolds, and more generally semi-Riemannian manifolds. In fact, this is closely related to noncompactness, as illustrated by our key example: the cylinder. In string theory, one thinks of the cylinder $\bR \times \bT$ as the worldsheet of a string, where the noncompact direction represents the time-axis. In order to give meaning to notions such as time, a causal structure is needed and, therefore, an indefinite metric.

Following Strohmaier \cite{99}, we start by setting up a general theory of Lorentz\-ian manifolds in terms of spectral triples. Then we return to the (Lorentzian) cylinder and study its isospectral deformation. For an introduction to semi-Riemannian and Lorentzian geometry, we refer to \dcite{beem} and \dcite{oneill}.

The description of a Lorentzian manifold in terms of spectral data requires a more general approach than that of Riemannian manifolds. This is enforced by the fact that the Lorentzian Dirac operator is no longer a self-adjoint operator on the Hilbert space of square integrable sections of the spin bundle. Rather it is Krein-self-adjoint on the Krein space of square integrable sections. Furthermore, the signature of the Lorentzian metric implies nonellipticity of the Dirac operator as a pseudodifferential operator acting on smooth sections. Before we go into details on this, we summarize some definitions concerning Krein spaces. For a more comprehensive overview we refer to \dcite{bognar} or to the lecture notes by Dritschel and Rovnyak, \dcite{100}.

Let $V$ be a nondegenerate indefinite inner product space. It is called \textbf{decomposable} if there are subspaces $V^-$, $V^+$ with $V = V^- \oplus V^+$ such that the inner product $(\cdot,\cdot)$ is negative definite on $V^-$ and positive definite on $V^+$. The inner product then defines a norm on these subspaces. If $V^-$ and $V^+$ are complete in these norms, then $V$ is called a \textbf{Krein space}. To every decomposition $V = V^- \oplus V^+$, we can associate an operator $J= -\textrm{id} \oplus \textrm{id}$, called \textbf{a fundamental symmetry}. This operator defines a positive definite inner product on $V$ by $\langle\cdot,\cdot\rangle_J:= (\cdot,J\cdot)$ which makes $V$ a Hilbert space. 
\begin{ex}
Consider flat Minkowski space, $V= \bR^n$, with inner product defined by $(x,y) = -x_0 y_0 + x_1 y_1 + \ldots + x_{n-1} y_{n-1}$. We have $V= \bR \oplus \bR^{n-1}$ and $J = \mathrm{diag}(-1, 1, \ldots, 1)$. Clearly, $\langle x, y \rangle_J = \sum_i x_i y_i$ is positive definite. 
\end{ex}
\subsection{Lorentzian spin geometry} \label{krein}
Our starting point will be an $n$-dimensional spin manifold $M$, equipped with a Lorentzian metric $g$, {i.e.} a metric with signature $(n-1,1)$. Spinors on this space-time are smooth sections of the spin bundle $S \to M$. In the following, we denote by $\gamma^\mu$ the curved gamma-matrices, whereas $\gamma^a$ are the flat ones \cite{lawson}. The flat gamma matrix $\gamma^0$ plays a special role in that it defines an operator $J:= \ii \gamma^0$ satisfying $J^2=1$. In fact, this operator is a fundamental symmetry of the space $L^2(M,S)$ of square integrable sections of the spin bundle. The space $L^2(M,S)$ is a Krein space endowed with the indefinite inner product
\bd
(\psi, \phi ) := \int_M \sum_{i,j} \psi_i^\ast(x) J_{ij} \phi_j(x) \sqrt{| g|} \dd x.
\ed
\subsubsection{Operators in Krein spaces}
In what follows we will make a clear distinction between self-adjoint operators and 
Krein-self-adjoint operators in $\cH$.

The Krein adjoint $A^\kast$ of a densely defined operator $A$ on a Krein space $\cH$ is defined with respect to the indefinite inner product $(\cdot,\cdot)$ on $\cH$.
One shows that $A^\kast = J A^\ast J$ and that $A$ is Krein-self-adjoint if and only if $AJ$ (or $JA$) is self-adjoint. See \dcite{99} for more details. 
According to Proposition 4.1 therein, we can formally write the square of a Krein-self-adjoint operator $A$ as
\be
(A)_J := \frac{1}{2} (A A^\ast + A^\ast A ). 
\ee
The $J$-dependency of this operator appears in the conjugation $^\ast$. It follows that the operator $(A)_J$ is self-adjoint and commutes with $J$. Hence, it is Krein-self-adjoint by the above remarks. 
\subsubsection{Spacelike reflections and fundamental symmetries}
Spacelike reflections make it possible to introduce a positive definite metric on a Lorentzian manifold (or, more generally, on a semi-Riemannian manifold). We give some of the basic notions and refer to \dcite{99} for a more detailed description. 
\begin{defn} \label{reflection}
Let $(M,g)$ be a semi-Riemannian manifold. A \textbf{spacelike reflection} $r$ is an automorphism of the vector bundle $TM$, such that
\begin{enumerate}
\item $g(r., r.) = g(.,.)$,
\item $r^2 = \id$,
\item $g^r(.,.) := g(.,r.)$ is a positive definite metric on $TM$. 
\end{enumerate}
This map induces a splitting of $TM$ in a direct sum $F_1 \oplus F_2$, such that 
\bd
r(x, k_1 \oplus k_2) = (x, -k_1 \oplus k_2).
\ed
The metric $g^r$ is called the \textbf{Riemannian metric associated to $r$}.
\end{defn}

If $M$ is a semi-Riemannian spin manifold, we can associate an operator $J_r$ to a spacelike reflection $r$. Let $e_0, e_1, \ldots, e_k$ be a local oriented orthonormal frame for $F_1$. We define $J_r := \ii^{k(k+1)} \gamma(e_0) \gamma(e_1) \cdots \gamma(e_{k-1})$. In the case of a Lorentzian manifold, we have $J_r = \ii \gamma^0$, which is a fundamental symmetry of the Krein space $L^2(M,S)$ as noted before.

\subsubsection{The Dirac operator}
We define the Dirac operator for a spin bundle $S \to M$ in local coordinates by
\be \label{diracop}
\Dirac := \gamma^\mu \nabla^S_\mu = \gamma^a e_a^\mu \nabla^S_\mu
\ee 
acting on smooth sections $\Gamma^\infty(S)$. Here, $\nabla^S$ is the lift of the Levi-Civita connection to the spin bundle. Its principal symbol $\sigma (\Dirac)$ satisfies the relation
\be
\sigma (\Dirac) (\xi)^2 = g(\xi, \xi).
\ee
This shows that the Dirac operator on a Lorentzian manifold is a nonelliptic pseudodifferential operator that is not self-adjoint. However, from the fact that $\ii J \mb \Dirac$ is self-adjoint, it follows that $D= \ii \Dirac$ is Krein-self-adjoint.

As ellipticity was an important property of the Dirac operator on a Riemannian manifold, we define an elliptic self-adjoint operator $\Delta_J$ using results of the previous subsection:
\be
\Delta_J := ((D)_J +1)^{1/2},
\ee
Ellipticity of this operator follows from considering its principal symbol:
\be \label{symbolD}
\sigma ( \Delta_J )(\xi) = \sqrt{g^r (\xi, \xi) }
\ee
where $ g^r $ is the Riemannian metric associated to $ g$. Furthermore, $\Delta_J$ is a pseudodifferential operator of order 1. This motivates the following definition. 
\begin{defn}
An \textbf{$n^+$-summable semi-Riemannian spectral triple} $\triple$ is given by an involutive algebra of operators $\cA$ in a Krein space $\cH$, such that $a^\ast = a^\kast$, and by a Krein-self-adjoint operator $D= D^\kast$ in $\cH$ such that
\begin{enumerate}
\item The commutators $[D, a] := Da -aD $ are bounded for any $a \in \cA$,
\item The operator $a \Delta_J^{-n}$ is in $\cL^{(1, \infty)}$, for all $a \in \cA$.
\end{enumerate}
Similar to Definition \ref{TripleNonUn}, the triple is called even if there is a grading operator $\chi$ on $\cH$ that satisfies the relations stated there with the only adjustification that now $\chi^\kast= \pm \chi$. 
\end{defn}
Of course, the triple $(C_0^\infty (M), L^2(M,S), D)$, where $D= \ii \Dirac$, is a semi-Riemannian spectral triple, called the \textbf{canonical triple} associated to the Lorentzian spin manifold $M$. If the dimension $n$ of $M$ is even, there is a $\bZ_2$-grading on the Hilbert space given by $\chi = \ii^{n/2} \gamma^0 \cdots \gamma^{n-1}$ so that the canonical triple is even. If the dimension of $M$ is odd, the canonical triple is odd. Since the self-adjoint operator $\Delta_J$ associated to $D$ is an elliptic pseudodifferential operator of order 1 as noted before, we have the following. Compare with Proposition \ref{connestrace}. 
\begin{prop}
Let $f$ be an integrable function on an $n$-dimensional Lorentzian manifold $M$, then
\bd
\int_M f(x) \sqrt{| g|} \dd x = \frac{n(2\pi)^n}{2^{[n/2]}\Omega_n} \trace_\omega(f \Delta_J^{-n})
\ed
\end{prop}
For the canonical triple, the fundamental symmetries of the form $J_r$ for some spacelike reflection $r$ play an important role. The analogue of such fundamental symmetries in the general case is given by the admissible fundamental symmetries as were defined in \dcite{99}. Therein, it is proved that the admissible fundamental symmetries of the canonical triple, are indeed exactly those of the form $J_r$. 
\subsection{Hochschild cocycles}
We associate a Hochschild $n$-cocycle to the canonical triple $(C^\infty_0 (M),\mb L^2(M,S),\mb D)$ as follows:
\be
\psi_D^\omega (a_0, a_1, \cdots, a_n) = \trace_\omega(a_0 [D,a_1 ]\cdots[D, a_n] |\Delta_J|^{-n} ).
\ee
Another cocycle can be constructed using the following result.
\begin{thm}
Let $(C^\infty_0 (M),\mb L^2(M,S),\mb D)$ be the semi-Riemannian canonical triple as defined before. Then $(C^\infty_0 (M),\mb L^2(M,S),\mb \Delta_J)$ is an $n^+$-summable spectral triple. 
\end{thm}
\textit{Proof. }
The only nontrivial condition to prove is the boundedness of $[\Delta_J, f]$ for any $f\in C^\infty_0(M)$. Since $\Delta_J$ is a pseudodifferential operator of order $1$, $[\Delta_J, f]$ is of order $0$, hence it is bounded. $\Box$

We define the following Hochschild $n$-cocycle on $C_0^\infty(M)$
\be
\psi_{\Delta_J}^\omega (a_0, a_1, \cdots, a_n) = \trace_\omega (a_0 [\Delta_J, a_1] \cdots [\Delta_J, a_n] |\Delta_J|^{-n}).
\ee
Obviously the two Hochschild cocycles do not coincide. We illustrate this by the following example. 
\begin{ex}
Let $M$ be a compact two-dimensional manifold, equipped with a Minkowski metric. 
We evaluate the 2-cocycles $\psi_{\Delta_J}^\omega$ and $\psi_D^\omega$ using symbol calculus:
\bd
\psi_{\Delta_J}^\omega (f,g,h) = C \int_M f \mb \dd g \wedge *(\dd h) 
\ed
in contrast to
\bd
\psi_D^\omega (f,g,h)= C' \int_M f \mb \dd g \wedge \dd h
\ed
for some integration constants $C, C'$. Note the appearance of a Polyakov type functional for $\psi_D^\omega$ in this special case.
\end{ex}
\subsection{Isospectral deformation of the Lorentzian cylinder}
With the theory of semi-Riemannian spectral triples developed in the previous sections in our hands, we are now in a position to describe the geometry of the cylinder equipped with a semi-Riemannian metric. In this section, we will discuss the semi-Riemannian spectral triple that describes the cylinder equipped with a Minkowski metric. Then we discuss isospectral deformation in this case, similar to what has been done before in the Euclidean setting. Finally, we show that the set of admissible fundamental symmetries of the noncommutative cylinder coincides with the set of fundamental symmetries coming from spacelike reflections in spinor space. 

The cylinder can be described by the following semi-Riemannian spectral triple:
\begin{eqnarray*}
\cA &:=& C^\infty_0(\bR^n \times \bT^d  );\\
\cH &:=& L^2 (\bR^n \times \bT^d ) \otimes \bC^{2^{[(n+d)/2]} };\\
D &:=& \ii \mb \Dirac.
\end{eqnarray*}
Here $\Dirac = \gamma^a \partial_a$, where the gamma-matrices satisfy $\{ \gamma^a, \gamma^b \} = 2 \eta^{ab}$ for the flat (Minkowski) metric $\eta= (-1, 1, \ldots, 1)$. 

In order to obtain a noncommutative manifold, we perform isospectral deformation of the Lorentzian cylinder, along the same lines as we did before for the Euclidean cylinder. We restrict to $\cS(\bR^n \times \bT^d)$ in order to represent the deformed algebra $ \cC_{\hbar\theta}^{(n,d)}$ on the Hilbert space $\cH$. 
\begin{thm}
The triple $\big( \cC_{\hbar\theta}^{(n,d)}, \mb\cH, \mb D \big)$ is a semi-Riemannian spectral triple, which is an isospectral deformation of $(\cS(\bR^n \times \bT^d), \mb\cH, \mb D)$. 
\end{thm}
One could very well imitate the construction of the Riemannian spin geometry on $\cC^{2d}_\hbar$ to obtain a Lorentzian spin geometry on the noncommutative cylinder. However, it turns out that in order to obtain for example the Lorentzian distance function from a canonical triple, one needs a different approach\cite{102}.

An admissible fundamental symmetry $J$ for the triple $(\cS(\bR^n \times \bT^d), \mb\cH, \mb D)$ is also admissible for the noncommutative cylinder. Indeed, $\cC_{\hbar\theta}^{(n,d)} $ is invariant under conjugation with $J$. Invariance of $\pi(\Omega \cC_{\hbar\theta}^{(n,d)} )$ follows from the following lemma.
\begin{lma} \label{forms}
For the semi-Riemannian spectral triple $\big( \cC_{\hbar\theta}^{(n,d)}, \mb\cH, \mb D \big)$ we have
\bd
\pi \big( \Omega^p \cC_{\hbar\theta}^{(n,d)} \big) = \big\{ \sum_j a^j \gamma(v_1^j) \cdots \gamma(v_p^j) : \quad a^j \in \cC_{\hbar\theta}^{(n,d)}, v_i^j \in \bC^{2^{[(n+d)/2]} } \big \}.
\ed
\end{lma}
\textit{Proof.} Recall the theory of universal forms on nonunital algebras described before. Since $[D, a] = \ii \gamma(\bdd a)$ still holds, we have for $a_i \in \cC_{\hbar\theta}^{(n,d)}$,
\bd
\pi (\tilde{a}_0 \delta a_1 \cdots \delta a_p )= \pi(\tilde{a}_0) \pi(\partial_{\mu_1} a_1) \cdots \pi(\partial_{\mu_p} a_p) \mb \gamma(\dd x^\mu_1) \cdots \gamma(\dd x^\mu_p),
\ed
which is of the required form. For $p=0$ we have $\pi \big( \Omega^0 \cC_{\hbar\theta}^{(n,d)} \big) = \cC_{\hbar\theta}^{(n,d)}$.
\begin{flushright}
$\Box$
\end{flushright}
In section \ref{krein}, we saw that the set of admissible fundamental symmetries of the semi-Riemannian canonical triple, coincides with the set of fundamental symmetries coming from spacelike reflections in $\bC^{2^{[(n+d)/2]} }$. Strohmaier \cite{99} showed that this statement also holds for noncommutative tori with trivial centre. For the noncommutative cylinder, we restrict to the class described in section \ref{PropNCC}, where $n=d$. There, we proved the following isomorphism of $C^\ast$-algebras:
\be \label{isom}
C^{2d}_\hbar \cong \cB_0 (L^2 (\bT^d) ) \otimes C(\bT^d).
\ee
The appearance of the set of compact operators in the tensor product plays a central role in the following result. It turns out to hold in a slightly more general setting, {i.e.} in the case of a semi-Riemannian Dirac operator.
\begin{thm}
The set of admissible fundamental symmetries of the noncommutative semi-Riemannian cylinder $\big( \cC^{2d}_\hbar, \mb \cH, \mb D \big)$ coincides with the set
\bd
\big\{ J_r : r \mbox{ is a spacelike reflection of } \bC^{2^d} \big\}.
\ed
\end{thm}
\textit{Proof. }
Let $r$ be a spacelike reflection of $\bC^{2^d}$. It follows from Lemma \ref{forms} that $J_r$ is admissible. For the proof of the converse statement, suppose that $J$ is an admissible fundamental symmetry of $ ( \cA, \mb \cH, \mb D )$, where we take $\cA:= \cC^{2d}_\hbar$. Since $J$ commutes with all elements in $\cA$, we have
\bd
J \in \cA' \otimes \End(\bC^{2^d}),
\ed
where $\cA'$ is the commutant of $\cA$ in $\cB(L^2(\bR^d \times \bT^d))$. Since the opposite algebra $\cA^\circ$, which acts on $\cH$ from the right, is a dense subalgebra of $\cA'$, we have for its unitization, with formula (10.82) in \dcite{varilly}, 
\begin{eqnarray*}
\widetilde{ \cA}^\circ &=& \big\{ T \in (\cA^\circ)'' : \quad T \in \Dom^\infty \delta \big\} \\
&=& \big\{ T \in  \cA' : \quad T \in \Dom^\infty \delta \big\},
\end{eqnarray*}
where $\Dom^\infty \delta$ is the smooth domain of the derivation $\delta := [\Delta_J, . ]$. Here we used the fact that $\cA' = \widetilde{\cA}'$. Since $J$ is smooth, $J \in \Dom^\infty \delta$, and it follows that $J \in \widetilde{ \cA}^\circ \otimes \End(\bC^{2^d})$. Note that the construction in \dcite{varilly} relies on finitely generated projective modules so that it does not directly apply to nonunital algebras. 

Since $\pi (\Omega^p \cA)$ is invariant under conjugation with $J$, we have
\bd
[J \pi (\Omega^1 \cA) J, \widetilde{ \cA}^\circ ] = [\pi (\Omega^1 \cA), \widetilde{ \cA}^\circ ] = 0 .
\ed
In particular, for any $\tilde{a} \gamma(v) \in \pi (\Omega^1 \cA)$, we have $\tilde{a} [ J \gamma(v) J, \widetilde{\cA}^\circ ] = 0$ so that $ J \gamma(v) J$ has entries in the center of $\widetilde{\cA}^\circ$. Since $C(\cA)=C(C_\hbar^{2d})=0$, as can be seen from formula (\ref{isom}), we infer that $C(\widetilde{\cA} ) = \bC$. Hence, $ J \gamma(v) J$ must be an element of $\End(\bC^{2^d})$, so that $ J\gamma(v) J =  -\gamma(rv)$ for some endomorphism $r$ of $\bC^{2^d}$. One checks the conditions of Definition \ref{reflection} to conclude that $r$ is a spacelike reflection. Hence there exists $J_r$ such that $J\gamma(v) J= J_r \gamma(v) J_r$. Define the operator $a := J J_r$. It commutes with all $\gamma(v)$, so that $a \in \widetilde{\cA}^\circ$. Since $J_r \in \End(\bC^{2^d})$, $[a, J_r]=0$. One shows that $a^2 =a^\kast a = a a^\kast = 1$ and that
\bd
\langle \xi, a \xi \rangle_{J_r} = ( \xi, J_r a \xi ) = (\xi, J\xi) \geq 0, \quad (\xi \in \cH).
\ed
We conclude that $a=1$ and $J=J_r$. $\Box$

\section*{Acknowledgements}
I am grateful to N.P. Landsman for his support, assistance and inspirational thoughts. I would like to thank G.G.A. B\"auerle for providing the physical motivation for the presented ideas. I want to thank J. Brodzki, J. Cuntz, V. Gayral, E. Langmann, and A. Strohmaier for helpful comments and my colleagues at the KdV-Institute and at SISSA for several discussions.
\pagebreak
\appendix
\section{Morita equivalence of $\cC^{2d}_\hbar$ and $\cS(\bZ^d)$} \label{morita}
\begin{thm}
The Fr\'echet algebras $\cC^{2d}_\hbar$ and $\cS(\bZ^d)$ are Morita equivalent via the Fr\'echet bimodule $\cS(\bR^d)$, {i.e.} 
\bea \nonumber
\cS(\bR^d) \tensor[\cS(\bZ^d)] \cS(\bR^d) &\cong& \cC^{2d}_\hbar,\\ \nonumber
\cS(\bR^d) \tensor[\cC^{2d}_\hbar] \cS(\bR^d) &\cong& \cS(\bZ^d). 
\eea
Here $\tensor[\cA]$ denotes the completion of the tensor product over a Fr\'echet algebra $\cA$ in the projective tensor product topology. 
\end{thm}
For notational convenience, we restrict to the case $d=1$. Recall that $\cC^{2}_\hbar$ is the Fr\'echet algebra $\cS(\bR \times \bT)$ equipped with the following product
\bd
(F \ast G ) \mbox (x,t)=\int_\bR \dd y F(y, t) G(x-y, t-\pi(y))
\ed
where $\pi : \bR \to \bR / \bZ \cong \bT $ is the natural projection.
We equip it with the following submultiplicative seminorms\cite{113}
\bd
p^{\alpha,\beta,\gamma} (F) = \int_\bT \dd t \int_\bR \dd x (1+|x|)^\alpha |\partial_x^{\beta} \partial_t^\gamma F(x ,t)|.
\ed
The algebra $\cS(\bZ)$ is equipped with the usual convolution product and the corresponding submultiplicative seminorms
\bd
q^\alpha (a) = \sum_{n \in \bZ} (1+|n|)^\alpha |a(n)|.
\ed
\textit{Proof.}
The module $\cS(\bR)$ is a Fr\'echet $\cC^{2}_\hbar-\cS(\bZ)$ bimodule in the following sense. First of all, it consists of differentiable functions on $\bR$ with the topology given by the seminorms $\nu^{\alpha,\beta}$
\be
\nu^{\alpha,\beta} (f) := \int_\bR \dd x (1+|x|)^\alpha | \partial_x^\beta f(x)|
\ee
The left and right actions of $\cC^{2}_\hbar$ and $\cS(\bZ)$ are defined by
\begin{eqnarray}
F \cdot f (x) &=& \int_\bR \dd y F(x-y, \pi(x) ) f(y),\\
f \cdot a(x) &=& \sum_n a(n) f(x+n).
\end{eqnarray}
One checks that these actions are continuous and that $(F \ast G) \cdot f = F \cdot (G \cdot f)$, $f \cdot (a \ast b)=(f \cdot a) \cdot b$. Furthermore, compatibility of both actions, $(F \cdot f) \cdot a = F \cdot (f \cdot a)$ is easily checked. We write $_{\cC^{2}_\hbar} \cS(\bR)_{\cS(\bZ)}$.
\\
We will endow $\cS(\bR)$ also with the structure of a $\cS(\bZ)-\cC^{2}_\hbar$ bimodule:
\begin{eqnarray}
f \cdot F (x) &=& \int_\bR \dd y F(y-x, \pi(y) ) f(y),\\
a \cdot f(x) &=&\sum_n a(n) f(x-n).
\end{eqnarray}
Again, this module satisfies the right properties and we write $_{\cS(\bZ)} \cS(\bR) _{\cC^{2}_\hbar}$.
\\[3mm]
Recall that an essential Fr\'echet $A$-modules $X$ satisfies $A \cdot X \subset X$ densely \cite{doran}.
\begin{lma}
The Fr\'echet bimodules $_{\cC^{2}_\hbar} \cS(\bR)_{\cS(\bZ)}$ and $_{\cS(\bZ)} \cS(\bR) _{\cC^{2}_\hbar}$ are essential.
\end{lma}
\textit{Proof. } 
Since the algebra $\cS(\bZ)$ is unital for the convolution product, there is nothing to prove there. The algebra ${\cC^{2}_\hbar}$ has an approximate identity $\{e_\lambda\}_{\lambda \in \Lambda}$ defined by
\bd
e_\lambda (x,t) := \frac{\lambda}{\sqrt{\pi}}e^{-\lambda x^2}.
\ed
We have $e_\lambda \cdot f \to f$ and $f \cdot e_\lambda \to f$ for $f \in \cS(\bR)$. $\Box$
\\[5mm]
We proceed by defining bilinear maps,
\bea
\widetilde{\phi} : \cS(\bR) \times \cS(\bR) &\to& \cC^{2}_\hbar \\ \nonumber
\widetilde{\phi} (f , g) (x,t) &=& \sum_n f(t-n) g(t-x-n)
\eea
\bea
\widetilde{\psi} : \cS(\bR) \times \cS(\bR) &\to& \cS(\bZ) \\ \nonumber
\widetilde{\psi} (f ,g) (n) &=& \int_\bR \dd x f(x) g(x -n)
\eea
One checks that the maps $\widetilde{\phi}$ and $\widetilde{\psi}$ are bounded bilinear module maps. They are balanced since one easily computes
\bea
\widetilde{\phi}(f \cdot a,g) &=& \widetilde{\phi}(f, a \cdot g), \\ \nonumber
\widetilde{\psi}(f \cdot F,g) &=& \widetilde{\psi}(f, F \cdot g).
\eea
Therefore, we can extend $\widetilde{\phi}$ and $\widetilde{\psi}$ to the tensor product:
\bea
\phi : \cS(\bR) \tensor[\cS(\bZ)] \cS(\bR) &\to& \cC^{2}_\hbar \\ 
\psi : \cS(\bR) \tensor[\cC^{2}_\hbar] \cS(\bR) &\to& \cS(\bZ)  \nonumber
\eea
Note that the maps $\widetilde{\phi}$ and $\widetilde{\psi}$ satisfy the following properties
\bea \label{couple}
\widetilde{\phi}(f,g) \cdot h &=& f \cdot \widetilde{\psi}(g,h),\\ \nonumber
\widetilde{\psi}(f,g) \cdot h &=& f \cdot \widetilde{\phi}(g,h).
\eea
\begin{lma}
The module morphisms $\phi$ and $\psi$ are surjective. 
\end{lma}
\textit{Proof. }
Let $F \in \cC^{2}_\hbar$. Define $H \in \cS(\bR) \tensor[\cS(\bZ)] \cS(\bR)$ by
\bd
H(x,y) := f(x) F(x-y,\pi(x)),
\ed
where $f \in \cS(\bR)$ satisfies $\sum_n f(t-n) =1 $ for all $t \in [0,1)$. Then $\phi (H) = F$. \\
For surjectivity of $\psi$ it is enough to construct a function $f \in \cS(\bR)$ with $\psi(f \otimes f) = \mathbf{1}_{\cS(\bZ)}$. Clearly, this holds for a function $f$ with $\mathrm{supp} f \in (0,1)$ and $\int \dd x | f(x) |^2 = 1$.$\Box$
\begin{lma}
The module morphisms $\phi$ and $\psi$ are injective
\end{lma}
\textit{Proof. }
Let $\sum_i \widetilde{\phi} (f_i, g_i) = 0$. Then
\bea
\widetilde{\phi} (f,g) \cdot \sum_i f_i \otimes_{\cS(\bZ)} g_i 
&=& \sum \widetilde{\phi}(f,g)\cdot f_i \otimes_{\cS(\bZ)} g_i \\ \nonumber
&=& \sum f \cdot \widetilde{\psi}(g, f_i) \otimes_{\cS(\bZ)} g_i \\ \nonumber
&=& \sum f \otimes_{\cS(\bZ)} \widetilde{\psi}(g, f_i) \cdot g_i \\ \nonumber
&=& f \otimes_{\cS(\bZ)} g \cdot \sum_i \widetilde{\phi} (f_i, g_i) \\ \nonumber
&=& 0,
\eea
using formula (\ref{couple}) twice. Hence, $F \cdot \sum_i f_i \otimes_{\cS(\bZ)} g_i =0$ for all $F \in \cC^{2}_\hbar$. Since $\cC^{2}_\hbar$ has an approximate identity it follows that $\sum_i f_i \otimes_{\cS(\bZ)} g_i =0 $. Since $\cS(\bZ)$ is unital, we find similarly injectivity of $\psi$. $\Box$
\\[5mm]
This completes the proof of the theorem. 

\pagebreak

\end{document}